\documentstyle[12pt]{article}

\begin{document}

\author{George Vlasov \\
Landau Institute for Theoretical Physics and \\
Moscow Aviation Institute, \\
Moscow, Russia\thanks{
E-mail: vs@itp.ac.ru}}
\title{Discontinuities on strings}
\maketitle

\begin{abstract}
Instead of the infinitesimal extrinsic and intrinsic perturbations on
strings, considered so far, we discuss the evolution and propagation of
finite-amplitude perturbations. Those intrinsic perturbations may result in
appearance of stable discontinuities similar to the shock waves.
\end{abstract}

\sloppy

For representing the cosmic strings \cite{Vilenkin85,Witten85} the simplest
possibility is a well-known Goto-Nambu model with the Lagrangian density $%
\Lambda =-m^2$. The energy density $U$ and tension $T$ have the
microscopically constant values $U=T=m^2$. The extensions \cite
{NO87,VV87,BPS88,Peter92,CSM94} are based on the introduction of the current
term $\chi =-h^{ab}\psi _{,a}\psi _{,b}$ - where $\psi _{,a}$ is a partial
derivative (with respect to 2-dimensional coordinates $\sigma ^a$ on the
world sheet) of a stream potential function $\psi $, on which the Lagrangian 
$\Lambda $ is dependent and $h^{ab}=g_{\mu \nu }\,x_{,a}^\mu \,x_{,b}^\nu $
is the induced surface metric. The energy-momentum tensor of the string has
the form \cite{Carter89a,Carter90} 
\begin{equation}
T^{\mu \nu }=Uu^\mu u^\nu -Tv^\mu v^\nu  \label{energmom}
\end{equation}
and obeys the conservation law 
\begin{equation}
\eta _\nu ^\rho \,\nabla _\rho \,T^{\mu \nu }=0  \label{consEM}
\end{equation}
where $u^\mu $ and $v^\mu $ are respectively the mutually orthogonal
time-like and space-like unit tangent vectors: $u^\mu u_\mu =-1=-v^\mu v_\mu
\,$, $u^\mu v_\mu =0$, and the fundamental tensor is 
\begin{equation}
\eta _\mu ^\nu \equiv e^{\nu \rho }e_{\rho \mu }=v^\nu v_\mu -u^\nu u_\mu
\qquad e^{\nu \rho }\equiv u^{[\nu }v^{\rho ]}  \label{fundtensor}
\end{equation}
These unit vectors correspond to the two conservation laws 
\begin{equation}
\eta _\mu ^\nu \nabla _\nu \,j^\mu =0\qquad \eta _\mu ^\nu \nabla _\nu
\,n^\mu =0  \label{conscur}
\end{equation}
for tangential currents $j^\mu =\mu \,v^\mu $ and $n^\mu =\nu \,u^\mu $,
where 
\begin{equation}
\mu =\exp \left\{ \int \frac{dT}{T-U}\right\}  \label{chempot}
\end{equation}
is the chemical potential or, the effective mass, and 
\begin{equation}
\nu =\exp \left\{ \int \frac{dU}{U-T}\right\}  \label{density}
\end{equation}
is the particle number density. The latter two quantities are related with
the energy density and the tension by the formula 
\begin{equation}
U=T+\mu \nu  \label{connection}
\end{equation}
The perturbations introduced in the equations of motion (\ref{consEM}), (\ref
{conscur}) may belong to the two classes \cite{Carter89b}, namely the
extrinsic perturbations of the world sheet (wiggles) and the sound type
longitudinal perturbations within the world sheet (jiggles) propagating with
the speed 
\begin{equation}
c_E^2=\frac TU  \label{soundextr}
\end{equation}
and 
\begin{equation}
c_L^2=\frac \nu \mu \frac{\partial \mu }{\partial \nu }=-\frac{\partial T}{%
\partial U}  \label{soundlong}
\end{equation}
respectively. The equation of state connecting the energy-density and
tension allows to find explicit results to express any quantity as a
function of argument $T$ or $U$ as well as of $\nu $ or $\mu $.

In the present paper we shall not restrict ourselves with the small
amplitude ''acoustic-type'' perturbations but rather try to seek the
possibility of discontinuities. The difference from the ''acoustic'' regime
discussed so far is that conditions and parameters ahead and behind the
front of discontinuity are not the same.

We use a natural system of units, the Plank and Boltzmann constants are
$\hbar =k_B=1$.

The appearance of discontinuities results from instability of the relevant
''acoustic'' perturbations. At this step the analysis is similar to that in
the case of usual waves \cite{LL87}. Consider an infinitesimal perturbation
of an arbitrary profile 
$U\left( x^\nu \right) =U_0+\Delta U\left( x^\nu\right)$ [or 
$\nu \left( x^\nu \right) =\nu _0+\Delta \nu \left( x^\nu\right)$], 
for instance a sinusoidal wave, which travels with the sound
speed $c_s$. If
\begin{equation}
\frac{dc_s}{d\nu }\neq 0  \label{neq}
\end{equation}
i.e. the perturbation propagates with velocity $c_s\left( \nu \right)$
depending on the perturbation magnitude $\nu$, the neighboring pieces of
the initial profile will run with different velocity:
\begin{equation}
c_s\left( x^\nu \right) =c_s|_{\nu =\nu _0}+\frac{\partial c_s}{\partial \nu
}\,\Delta \nu \left( x^\nu \right)  \label{diff}
\end{equation}
In the case of the sinusoid each bottom tends to overtake the adjacent top.
For instance, if $\partial c_s/ \partial \nu >0$, the pieces of the profile
corresponding to the largest density $\nu$ will run with the highest speed;
in turn, the higher the speed the larger the density; thereby, the tops run
faster than the bottoms and the profile must change. After all, the initial
smooth profile degenerates into an ambiguous solution, since the bifurcation
points take place; for example, a sinusoidal wave becomes a saw-tooth. This
behavior of the ''acoustic'' waves may pertain both to the extrinsic
(transversal) and intrinsic (longitudinal) perturbations on strings, in
other words, the ''jiggles'' or ''wiggles'' are those ''embryos'' which
finally can yield the catastrophes. The latter, in the case of a
longitudinal perturbation, implies no more than a discontinuity similar to
usual shock wave in continuous medium. From this process of the shock wave
formation, which accrue from the spontaneous acoustic wave growth, we may
easily predict the possible configuration of discontinuities: the sign of
derivative in (\ref{diff}) determines directly the sign of $\Delta \nu $, or
else, the sign of $\Delta \mu $ coincides with sigh of $dc_L/d\mu $. This
may be written as 
\begin{equation}
sign\left\{ \frac{dc_L}{d\nu }\right\} =sign\left\{ \Delta \nu \right\}
\qquad sign\left\{ \frac{dc_L}{d\mu }\right\} =sign\left\{ \Delta \mu
\right\}  \label{sign}
\end{equation}

On the other hand, one may expect that the growth of an extrinsic
perturbation leads to a collapse corresponding to the string
self-intersection \cite{KT82} appearance; the detailed non-estimate analysis
is beyond the scope of the present paper where we direct our attention to
the stationary solutions, however, we have not missed the ability to point
qualitatively this feasible scenario. For a sinusoidal wave \cite{BS96} with
a given frequency $\omega $ (or length $\lambda $) and finite magnitude 
$\Delta \nu _0$ (or $\Delta U_0$) the catastrophe occurs in a time interval 
\begin{equation}
\tau =\frac \lambda {\Delta c_s}=\frac{2\pi }\omega \frac{c_s}{\,\left(
dc_s/d\nu \right) \Delta \nu _0}=\frac{2\pi }\omega \frac{c_s}{\,\left(
dc_s/dU\right) \Delta U_0}  \label{time}
\end{equation}
which can be understood as a lifetime of perturbations. While the quantity 
$\Delta U_0$ 
is equal approximately to the effective temperature $\Theta $ 
\cite{Carter90,CSM94}. Since the characteristic frequency of the black-body
radiation is proportional to temperature: 
$\omega \sim \Theta$, 
the time (\ref{time}) is estimated as 
\begin{equation}
\tau \sim \frac 1{\Theta ^2}\frac{c_s}{\,\left( dc_s/dU\right) }
\label{time2}
\end{equation}
omitting dimensionless constant in the right side.

As for the longitudinal shock waves, constraint (\ref{neq}) allows only the
possibility of discontinuitis but does not provide their real existence.
That is why, the evolutionary condition is also necessary. Nevertheless we
can immediately exclude the strings with the simplest Lagrangians 
\begin{equation}
\Lambda =-m^2\qquad \Lambda =-m^2+\chi /2  \label{simple}
\end{equation}
from this category, since the relevant sound speed remains a constant. The
discontinuities are conceivable in the light of more realistic ''fourth''
and ''fifth'' models \cite{CP95} (already applied to recent investigations 
\cite{MS98}) which we shall use bellow.

Consider the longitudinal perturbations described by 
equations (\ref{conscur}) and the characteristic vector \cite{Carter89b} 

\begin{equation}
L_\rho =\frac{u_\rho \pm c_Lv_\rho }{\sqrt{1-c_L^2}}  \label{charact}
\end{equation}
Written in the form of discontinuities \cite{Taub78}, the conservation laws 
(\ref{conscur}) yield 

\begin{equation}
\left[ \eta _\mu ^\nu \lambda _\nu \,j^\mu \right] \equiv \eta _\mu ^\nu
\lambda _{+\nu }\,j_{+}^\mu -\eta _\mu ^\nu \lambda _{-\nu }\,j_{-}^\mu
=0\qquad \eta _\mu ^\nu \lambda _{+\nu }\,n_{+}^\mu -\eta _\mu ^\nu \lambda
_{-\nu }n_{-}^\mu =0  \label{d-cur}
\end{equation}
where the normal unit vector 
$\lambda_\rho$ 
is chosen in the form 

\begin{equation}
\lambda _\rho =\frac{u_\rho +wv_\rho }{\sqrt{1-w^2}}  \label{lambda}
\end{equation}
and the indices 
$+$ 
and 
$-$ 
correspond to the states at different
sides of the discontinuity. Substituting (\ref{fundtensor}) in (\ref{d-cur})
we find two equations 

\begin{equation}
\frac{w_{+}^2n_{+}^2}{w_{+}^2-1\ \,}=\frac{w_{-}^2n_{-}^2}{w_{-}^2-1}
\label{l1}
\end{equation}

\begin{equation}
\frac{\ \mu _{+}^2}{w_{+}^2-\ 1}=\frac{\ \mu _{-}^2}{w_{-}^2-1}  \label{l2}
\end{equation}
which determine the velocity of the discontinuity 

\begin{equation}
w_{-}^2=\frac{n_{+}^2}{\mu _{+}^2}\frac{\mu _{+}^2-\mu _{-}^2}{%
n_{+}^2-n_{-}^2}  \label{w-l}
\end{equation}
and the velocity behind the discontinuity 

\begin{equation}
w_{+}^2=\frac{n_{-}^2}{\mu _{-}^2}\frac{\mu _{+}^2-\mu _{-}^2}{%
n_{+}^2-n_{-}^2}  \label{w+l}
\end{equation}
Solutions (\ref{w-l}) and (\ref{w+l}) in the acoustic limit
$\Delta =n_{+}-n_{-}\rightarrow 0$
and
$\Delta \mu =\mu _{+}-\mu _{-}\rightarrow 0$, are reduced to

\begin{equation}
\frac{w_{-}^2}{c_{-}^2}=1+\frac 12\left[ 1-c_{-}^2+\frac{\mu _{-}}2\frac{%
\partial c_{-}^2}{\partial \mu _{-}}\right] \frac{\Delta \mu }{c_{-}^2}%
+O\left( \Delta \mu ^2\right)   \label{wd-l}
\end{equation}

\begin{equation}
\frac{w_{+}^2}{c_{+}^2}=1-\frac 12\left[ 1-c_{+}^2+\frac{\mu _{+}}2\frac{%
\partial c_{+}^2}{\partial \mu _{+}}\right] \frac{\Delta \mu }{c_{+}^2}%
+O\left( \Delta \mu ^2\right)  \label{wd+l}
\end{equation}
or 
\begin{equation}
\frac{w_{-}^2}{c_{-}^2}=1+\frac 12\left[ 1-c_{-}^2+\frac{n_{-}}{2c_{-}^2}%
\frac{\partial c_{-}^2}{\partial n_{-}}\right] \Delta n+O\left( \Delta
n^2\right)  \label{wd-l2}
\end{equation}

\begin{equation}
\frac{w_{+}^2}{c_{+}^2}=1-\frac 12\left[ 1-c_{+}^2+\frac{n_{+}}{2c_{+}^2}%
\frac{\partial c_{+}^2}{\partial n_{+}}\right] \Delta n+O\left( \Delta
n^2\right)   \label{wd+l2}
\end{equation}
where the sound speed ahead and behind the discontinuity, $c_{-}$ and $c_{+}$
respectively, is given through the relevant variables by the formula (\ref
{soundlong}).

The evolutionary condition of the discontinuity existence requires to be 
\cite{Thorne73,LL87} 
\begin{equation}
w_{-}>c_{-}\qquad w_{+}<c_{+}  \label{evol}
\end{equation}
Constraint (\ref{evol}), in the acoustic limit, in light of (\ref{wd-l}) and
(\ref{wd+l}), implies that 
\begin{equation}
M\Delta \mu =N\Delta n>0  \label{evol2}
\end{equation}
\begin{equation}
M=\left( 1-c_L^2+\frac \mu 2\frac{\partial c_L^2}{\partial \mu }\right)
\frac 1{c_L^2}  \label{mm}
\end{equation}
\begin{equation}
N=1-c_L^2+\frac \nu {2c_L^2}\frac{\partial c_L^2}{\partial \nu }  \label{nn}
\end{equation}
The longitudinal speed obtained for the ''fourth'' \cite{CP95} 
\begin{equation}
c_L^2=\frac{m_{\star }^2-3k_0^2\nu ^2}{m_{\star }^2+k_0^2\nu ^2}\qquad c_L^2=%
\frac{m_{\star }^2-k_0^2\mu ^2}{m_{\star }^2+3k_0^2\mu ^2}  \label{l-fourth}
\end{equation}
and ''fifth'' equation of state 
\begin{equation}
c_L^2=\frac{m_{*}^2-k_0^2\nu ^2}{m_{*}^2+k_0^2\nu ^2}\qquad c_L^2=\frac{%
m_{*}^2-k_0^2\mu ^2}{m_{*}^2+k_0^2\mu ^2}  \label{l-fifth}
\end{equation}
is expressed through $\nu$ or through $\mu$ in the ''magnetic'' and
electric ''regime'' respectively. Hence, 
\begin{equation}
\frac{\partial c_L^2}{\partial \nu }=-\frac{8k^2\nu m^2}{\left( m^2+k^2\nu
^2\right) ^2}<0\qquad \frac{\partial c_L^2}{\partial \mu }=-\frac{8k^2\mu m^2%
}{\left( m^2+3k^2\mu ^2\right) ^2}<0  \label{diff4}
\end{equation}
for the ''fourth'' model and 
\begin{equation}
\frac{\partial c_L^2}{\partial \nu }=-\frac{4k^2\nu m^2}{\left( m^2+k^2\nu
^2\right) ^2}<0\qquad \frac{\partial c_L^2}{\partial \mu }=-\frac{4k^2\mu m^2%
}{\left( m^2+3k^2\mu ^2\right) ^2}<0  \label{diff5}
\end{equation}
for the ''fifth'' model. Note that in the magnetic regime the variables are
expressed through $\nu$, while $\mu$ is used as the main argument in the
electric regime.

Substitution of Eqs. (\ref{diff4}) and (\ref{diff5}) in Eqs. (\ref{mm}) and (%
\ref{nn}) yields 
\begin{equation}
N=-M=-\frac 34\frac{\left( 1-c_L^2\right) ^2}{c_L^2}  \label{fourth}
\end{equation}
for the ''fourth'' model, while 
\begin{equation}
N=-M=-\frac 12\frac{\left( 1-c_L^2\right) ^2}{c_L^2}  \label{fifth}
\end{equation}
for the ''fifth'' model. Taking into account the equations of state \cite
{CP95} which connects $\mu$ with $n$ and formulae (\ref{evol2})-(\ref{nn}),
(\ref{fourth}), (\ref{fifth}), we always find 
\begin{equation}
\Delta n<0\qquad \Delta \mu >0  \label{nm}
\end{equation}
\begin{equation}
\Delta T<0\qquad \Delta U>0  \label{tu}
\end{equation}
in both ''magnetic'' and ''electric'' regimes, i.e. without regard of the
current term $\chi$ sign.

Do real shock waves satisfy (\ref{nm}), (\ref{tu})? From the above analysis
of the spontaneous acoustic wave growth, which leads to the shock wave
formation, we may predict the possible configuration of discontinuities: the
sign of derivative in (\ref{diff}) determines directly the sign of 
$\Delta n$; or else the sign of 
$\Delta \mu$ coincides with the sign of 
$dc_L/d\mu $. Hence, according 
to (\ref{diff4}) and (\ref{diff5}), $\Delta n<0$\ in the
magnetic regime, while $\Delta \mu <0$ in the electric regime. 
The latter
possibility is forbidden by constraint (\ref{nm}). Thereby, the shock waves
are permitted only in magnetic regime, i.e. in the case of time-like
currents.

The evolutionary condition (\ref{evol}), (\ref{nm}), (\ref{tu}) itself \cite
{Thorne73,LL87} is not sufficient for existence of longitudinal
discontinuities, since the latter (although, in the local sense, it can be
compared with a wave in one dimension which is not subject to that
instability) may be unstable with respect to the extrinsic disturbance of
the world sheet. This is an analogue of the so-called ''goffer-type''
instability \cite{LL87}, conceivable as a ruffle on the wave surface. The
instability takes place if \cite{Carter89b} 
\begin{equation}
\frac TU<0  \label{ruffle}
\end{equation}
Thus, if the energy density and tension is positive before the front, the
shock wave will be stable if 
\begin{equation}
U_{+}>0\qquad T_{+}>0  \label{behind}
\end{equation}
The latter constraint is automatically satisfied for the parameters in (\ref
{l-fourth})-( \ref{l-fifth}) varying in the allowed range \cite{CP95}.

In future the present study may be extended to e.g. rotating string loops or
rings (known also as 'vortons') and, I can suspect, the recognition of
phenomenon similar to the elastic wave inertia in a rotating ring \cite
{Bryan890,Zhuravlev93}.

\end{document}